\documentclass[twocolumn, lengthcheck]{revtex4-1}
\usepackage{amsmath,amssymb}
\usepackage{epsfig}
\usepackage{graphicx}
\usepackage{amsmath}
\usepackage{slashed}
\usepackage{amsfonts}
\usepackage{epstopdf}
\usepackage{color}

\newcommand{\be}{\begin{equation}}
\newcommand{\ee}{\end{equation}}
\newcommand{\bear}{\begin{eqnarray}}
\newcommand{\eear}{\end{eqnarray}}
\newcommand{\ba}{\begin{array}}
\newcommand{\ea}{\end{array}}

\def\be{\begin{eqnarray}}
\def\ee{\end{eqnarray}}
\def\bea{\be}
\def\eea{\ee}

\def\roughly#1{\mathrel{\raise.3ex\hbox{$#1$\kern-.75em%
\lower1ex\hbox{$\sim$}}}}

\begin{document}

\title{Holographic tetraquarks and the newly observed $T_{cc}^{+}$ at LHCb}

\author{Yizhuang Liu}
\email{yizhuang.liu@uj.edu.pl}
\affiliation{Institute of Theoretical Physics, Jagiellonian University, 30-348 Kraków, Poland}

\author{Maciej A. Nowak}
\email{maciej.a.nowak@uj.edu.pl}
\affiliation{Institute
of Theoretical Physics and Mark Kac Center for Complex Systems Research,
Jagiellonian University, 30-348 Kraków, Poland}

\author{Ismail Zahed}
\email{ismail.zahed@stonybrook.edu}
\affiliation{Department of Physics and Astronomy, Stony Brook University, Stony Brook, New York 11794-3800, USA}


\date{\today}
\begin{abstract}
We  describe a heavy and exotic tetraquark state as a holographic molecule, by binding the lightest heavy-light meson $(0^-, 1^-)$ multiplet to a flavored sphaleron, in the bulk of the Witten-Sakai-Sugimoto model.   Bound tetraquark states emerge as Efimov states in the heavy quark limit, with a binding energy for a charm tetraquark comparable to the $T_{cc}^+$ recently reported by the LHCb collaboration,  but with a substantially smaller width, for a large but finite $^\prime$t Hooft coupling. Fixing the  parameters  of the model at the empirical mass of $T_{cc}^+$, allows for a prediction of  the bindings of the  undiscovered 
bottom-charm and bottom tetraquarks.
\end{abstract}

\maketitle

\setcounter{footnote}{0}

\section{Introduction}

Hadrons composed of heavy  ($Q$) and light ($q$)  quarks  have received considerable  interest lately, due to the flurry of results stemming from electron and hadron colliders~\cite{BELLE,BESIII,DO,LHCb,LHCbx,LHCbxx}. These hadrons embody in  a remarkable way some key aspects of QCD:
the spontaneous breaking of chiral symmetry  for the light quarks, and a heavy quark spin flip symmetry~\cite{ISGUR,MACIEK2}. In the heavy quark mass limit, a heavy hadron with spin up is degenerate with its counterpart with spin down, and the resulting doublets with even and odd parity
are chiral partners of each other~\cite{MACIEK,BARDEEN}. The result is the chiral doubling phenomenon first observed by the Babar collaboration~\cite{BABAR}, and then confirmed by the CLEO collaboration~\cite{CLEOII}. Chiral doubling of heavy-light hadrons is likely  more remarkable with b-quarks at LHCb and BESIII.

The many discoveries by several  collaborations have led to renewed interest in
exotic heavy-light hadrons.
Spectacular discoveries are
  the $\chi_{c1}(3872)$, directly at the $D^{*0}$  and $\bar{D^0}$ threshold (minimal quark content $c\bar{c}q\bar{q}$), the family of Z states  $ Z_c(3900,4020,4050,4200,4430)$  (minimal quark content $c\bar{c}u\bar{d}$),  some of their strange cousins $Z_{cs}(3985,4000,4220)$, and  bottom analogues $ Z_b(10610,10650)$. These
  exotica  (tetraquarks) are thought to be
bound deuteron-like molecules of the type $(Q\bar q)(\bar Q q)$~\cite{MOLECULES,THORSSON,KARLINER,OTHERS,OTHERSX,OTHERSZ,OTHERSXX,LIUMOLECULE},
although alternative explanations have been presented~\cite{MACIEK2,RICHARD,MANOHAR,RISKA,SUHONG,MACIEK3}. In the molecular scenario, the chemical-like bonding is thought to be mediated mostly by pion-exchange, and perhaps at the origin of the newly discovered
and exotic charmed baryon-meson molecules of the type $(\bar Qq)(Qqq)$ such as the triplet
$P_c((4312),(4440),(4457))$ and the newly noticed  $P_c(4380)$, (although at the level of 3-4 $\sigma$ only), all with a minimal, pentaquark content $c\bar{c}uud$. Their strange cousin $P_{cs}(4459)$ has been also discovered.  Note that all these states have a hidden heavy flavor.

\begin{figure*}
{%
  \includegraphics[height=6cm,width=.49\linewidth]{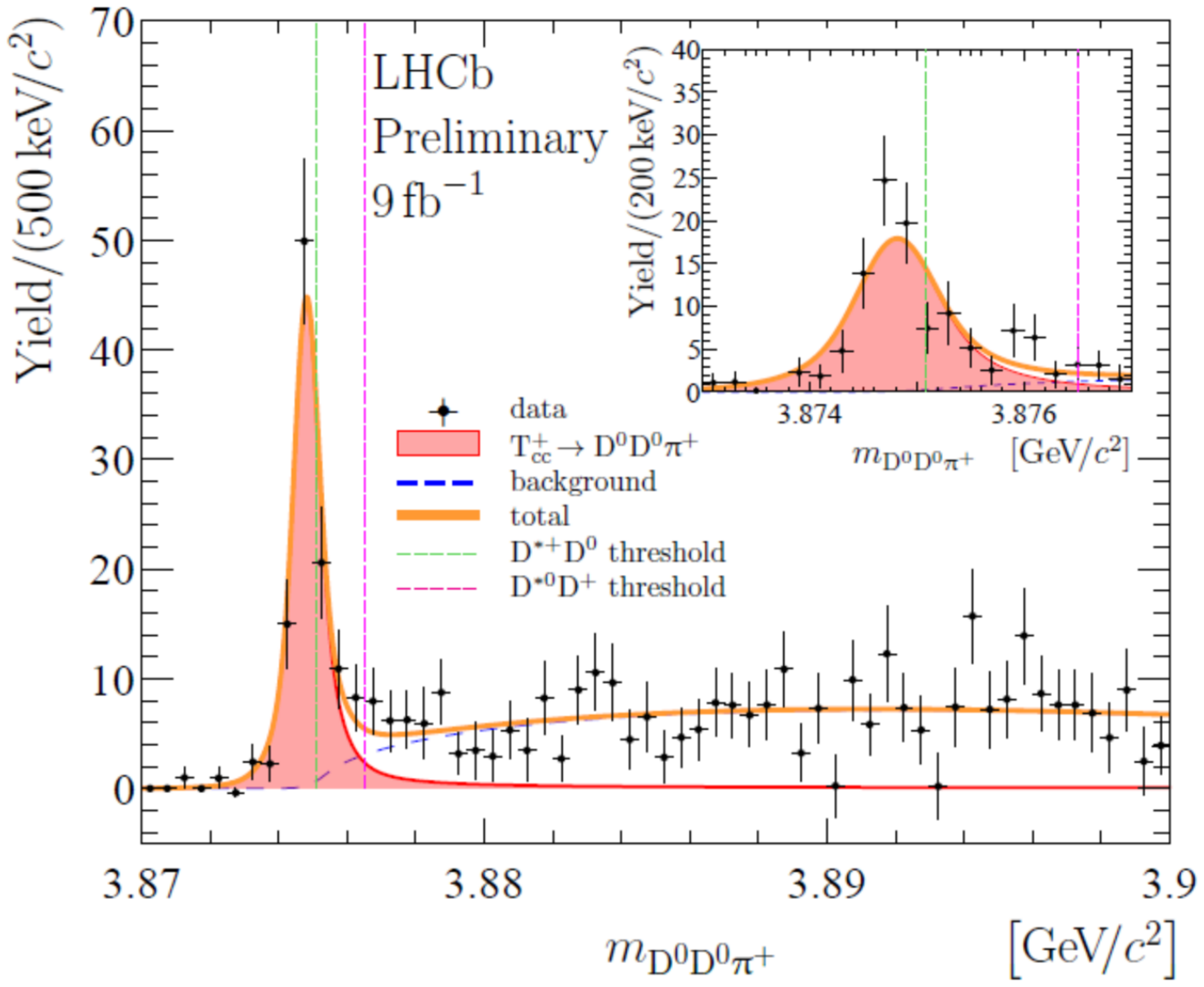}%
}
\caption{Newly measured  isoscalar charmed $T_{cc}^+$ tetraquark  by LHCb~\cite{LHCb:2021auc}.}
\label{fig-LHCB}
\end{figure*}

The recent discovery of doubly heavy baryons of the $ccq$  type suggests that an  approximate Savage-Wise symmetry~\cite{SW} may be at  work, where a heavy and compact diquark $QQ$ would be equivalent to a
heavy anti-quark $\bar Q$. This heavy diquark-anti-quark  {\bf supersymmetry} allows for mass relations not only between heavy baryons and mesons such as $\bar Q q$ and $QQq$, but also
between heavy baryons and tetraquarks  with hidden heavy flavor such as $Qqq$ and $\bar Q\bar Q qq$.
The recent quark model estimates for this last state are remarkable~\cite{MAREK,KR}. It is suggested that for
a compact $\bar b\bar b qq$ tetraquark, the binding energy is significant and about 200 MeV. If confirmed, this would be the first, non-molecular and truly exotic tetraquark state outside the standard quark model classification.

Exploratory lattice QCD simulations
appear to support the quark model prediction of a strongly bound b-tetraquark~\cite{MALT}. Given the difficulty to
analyze QCD in the confining regime, it is not easy to identify the mechanism at work in the formation of these exotics.
In recent years, holography has proven to be a useful framework for discussing QCD for a large number of colors $N_c$ and strong coupling
$\lambda$~\cite{HOLOXX,HOLOXXX,HOLOXXXX}. For  hadrons,
the formulation confines and breaks spontaneously chiral symmetry through geometry~\cite{SSX,SSXB,SSXBB,CSLIGHT,KOJI,CSTHREE}.
Its  extension to a heavy quark  exhibits explicit heavy-quark symmetry~\cite{LIUHEAVY} (for earlier approaches see~\cite{FEWX,CHIN}).
Light holographic baryons are instantons in bulk, while heavy holographic baryons are bound states of a heavy meson multiplet to the instanton. This mechanism is the dual
of  the Callan-Klebanov mechanism~\cite{CK} in the context of the Skyrme model~\cite{SKYRME} which we will review for clarity below.
For completeness, we  note that tetraquarks in the context of a holographic string construction have been discussed in~\cite{VENEZIANO,COBIX}, and using light cone holographic QCD in~\cite{BRODSKY}.

In this letter we revisit our recent holographic analysis of the charm and bottom tetraquark states~\cite{LIUNOWAK},  by refining their mass analysis and assessing their
strong decay widths. We take advantage of the fact, that the first open charm exotic state has just been discovered. Our  results are shown  to be compatible, with the newly reported  charmed tetraquark $T_{cc}^+$ by LHCb as shown in Fig.~\ref{fig-LHCB}, for a large value of the
$^\prime$t Hooft coupling. Using this experimental value,   more precise
estimate of the masses  are given for the mixed charm-bottom and bottom tetraquark states,  yet to be observed. We also provide generic arguments, why the width of the $T_{cc}^+$ is so narrow.

\section{Strange solitonic baryons}

In the large number of colors limit, QCD truncates to an effective theory of weakly coupled mesons where baryons are solitons.
The meson effective theory is chiefly chiral, consisting of the light mesons.
Once the effective mesonic description is fixed, the baryonic description follows with no new parameters. The soliton is usually characterized by a moduli following from the set of zero modes associated to the classical solution. The quantum numbers of the baryon follows by quantizing the moduli using the so-called collective coordinate method. This construction works well for two light flavors up-down, but when extended to strangeness, the method  fails phenomenologically.

Callan and Klebanov~\cite{CK} argued that the strange mass is somehow large, and therefore a strange quark  as a kaon cloud is more likely to bind to the soliton owing to its short Compton wavelength.  Specifically,  the fast vibrational modes (kaon) do not decouple from the slow rotational moduli (soliton), and generate an effective potential (non-Abelian Berry phase)  in the Born-Oppenheimer approximation. As a result, the spin of the rotating soliton is shifted by the isospin of the kaon. This construction fares better phenomenologically.

This construction has been extended to charm and bottom heavy baryons~\cite{SKYRMEHEAVY,PENTARHO}. The difference with strangeness though is that  the partners of the kaon, i.e. $(0^-,1^-)=(D(B),D^*(B))$, are  degenerate leading to a degenerate baryon multiplet $({\frac 12}^+, {\frac 32}^+)$, in the heavy quark limit.
Parity doubling  suggests a nearby baryon multiplet $({\frac 12}^-, {\frac 32}^-)$ by binding to  the chiral partners $(0^+, 1^+)=(\tilde{D}(B),\tilde{D}^*(B))$~\cite{NEAR}.

\section{Holographic light   baryons}

In holographic QCD, confinement and chiral symmetry breaking can be addressed simultaneously
for instance in the Witten-Sakai-Sugimoto model~\cite{SSX}. Confinement follows from a stack of colored D4-branes,  in the double limit of a large number of colors and strong coupling, with the string tension fixed by the apparent horizon. The spontaneous breaking of chiral symmetry arises from the geometrical fusion of a pair of flavored D8-D$\bar8$-branes in the probe approximation. The model with only two parameters-the brane tension $\kappa$ and the Kaluza-Klein  compactification scale $M_{KK}$-  is in remarkable agreement with phenomenology~\cite{HIDDEN}.

Holographic baryons are flavor valued instantons in the probe D8-D$\bar8$  branes. Their topological charge is identified with baryon  charge, and their quantization follows from the quantization of the instanton moduli in bulk. Most noteworthy is the fact that the instanton size or equivalently the baryon core is fixed by geometry or equivalently  the BPS condition, making it independent of the nature of the mesons retained and/or their derivatives thereby solving a key problem in the Skyrme model.

The quantum moduli for the flavored instanton is the standard $R^4\times R^4/Z_2$ (flat space)~\cite{SSX}. We focus on  $R^4/Z_2$ which corresponds to the size
and global flavor SU(2) orientations, and denote by $y_I=\rho a_I$ the
coordinates on $R^4/Z_2$, with the SU(2) orientations parametrized by $a_I$ subject to the
normalization $a_I^2=1$,  and  $\rho$ the instanton size.  The collective Hamiltonian in polar coordinates
on the $R^4/Z_2$ moduli for the light holographic baryon,  is~\cite{SSXB}

\begin{widetext}
\be
\label{H1}
{\bf H}_{k}=-\frac 1{2m_k}\left(\frac 1{\rho^{\frac 32}}\partial^2_\rho\,\rho^{\frac 32}
+\frac 1{\rho^2}(\nabla^2_{S^3} -2m_kQ(k))\right)+\frac 12 m_k\omega_k^2\rho^2
\ee
\end{widetext}
All scales are in units of the KK scale $M_{KK}$ which is set to 1.
The $k=1$ labels the instanton path with topological charge 1. The
inertial parameters are $m_{k=1}=16\pi^2 aN_c$, $\omega_{k=1}^2=\frac 16$. The charge
$Q(k=1)=N_c/(40\pi^2 a)$ with  $a=1/(216\pi^3)$, characterizes the U(1) topological self-repulsion
within the instanton. The first two contributions in (\ref{H1}) are the kinetic  Laplacian in $R^4$,
and the last harmonic contribution is the  gravitational attraction induced by the warped
 holographic direction. A detailed derivation of (\ref{H1}) can be found in~\cite{SSXB} (see Eq. 5.9)
 and will not be repeated here.

The eigenstates of (\ref{H1}) are $T_l(a)R_{ln}$, with $T_l(a)$ as
spherical harmonics on $S^3$ with $\nabla^2T_l=-l(l+2)T_l$.
Under SO(4)$\sim$ SU(2)$\times$SU(2) they are in the symmetric $(\frac l2, \frac l2)$ representations, with the
two SU(2) identified by the isometry $a_I\rightarrow V_La_IV_R$. The left factor is the isospin rotation and the right factor is the space rotation with quantum numbers $I=J=\frac l2$. For instance the proton with spin-up carries a wavefunction $R(a)\sim (a_1+ia_2)$ with a rotational tower
($M_1=8\pi^2\kappa$)

\be
M=M_1+\left(\frac{(l+1)^2}6+\frac 2{15}N_c^2\right)^{\frac 12}+\frac 2{\sqrt{6}}
\ee

\section{Holographic exotic baryons}

Recently, two of us extended the holographic approach to the description of heavy-light mesons and baryons with manifest chiral and heavy quark symmetry~\cite{LIUHEAVY}. Heavy baryons emerge by binding a 5-dimensional $(0^-, 1^-)$ spin-1 multiplet to the flavored instanton in bulk. In the heavy
mass limit, the spin-1 meson transmutes to a spin-$\frac 12$ zero mode, leading to a rich heavy
baryon spectrum including exotics,  thereby extending the Callan-Klebanov mechanism to holography.

More specifically, the instanton moduli described above is extended to include a spin-$\frac 12$ Grassmannian $\chi$ to account for the spin $1\rightarrow \frac 12$ transmutation following the binding. The ensuing moduli for the exotic baryonic molecule follows also from (\ref{H1}) with  the general charge

\begin{widetext}
\be
\label{H2}
Q(k)=\frac {N_c}{40\pi^2 a}
 \left(q(k)+\frac {\lambda}{m_H}\frac {5\alpha_0(k)}{432\pi}\frac {N_Q}{N_c}
+30\alpha_1(k)\frac{N_Q}{N_c}+5\alpha_2(k)\frac{N_Q^2}{N_c^2}\right)
\ee
\end{widetext}
with $N_Q=\chi^\dagger\chi$.
For the instanton  $q(1)=1$ (topological charge) and $\alpha_0(1)=0$ (self-dual).
$\alpha_1(1)=-\frac 18$ characterizes the magnetic interaction of the heavy multiplet to the instanton,
and $\alpha_2(1)=\frac 13$ captures the U(1) repulsion between the bound heavy mesons. $N_Q=1,2, ...$ counts the number of bound mesons. More details regarding the charge (\ref{H2}) for $k=1$ can be found in~\cite{LIUHEAVY} (last reference Eq. 40).

The binding of any number of  heavy mesons and anti-mesons follows from the substitution $N_Q\rightarrow N_Q-N_{\bar Q}$. In general, the isospin (${\bf I}$) and spin  (${\bf J}$) now  decouple,  with the identification~\cite{LIUHEAVY}

\be
\label{H3}
{\bf J}=-{\bf I}+\chi^\dagger_Q{\bf T}\chi_Q
\ee
The isospin-spin quantum numbers for the heavy exotic baryons are now shifted

\be
\label{H4}
IJ\equiv \left(\frac l2, \frac l2\right)\rightarrow \left(\frac l2, \frac l2 \bigoplus_{i=1}^{N_Q} \frac 12\right)
\ee

\section{Holographic heavy tetraquark}

The predicted tetraquark in the context of the quark model is  more challenging to describe using a topological molecular formulation since it is a boson and not a fermion. Here we propose to bind a heavy multiplet $(0^-,1^-)$ to a sphaleron path as a topological tetraquark molecule, in total correspondence with the heavy holographic baryons described above.  In the process quantum numbers get transmuted. This remarkable construction provides a topological realization for the Savage-Wise symmetry~\cite{SW} whereby a fermion is continuously deformed to a boson along the sphaleron hill, in the holographic dual approach.

With this in mind, we observe that the instanton as an O(4) gauge configuration belongs to a class of tunneling
paths with fixed Chern-Simons number, that cross the sphaleron hill, with the instanton at the bottom and the sphaleron at the top. These configurations  are given by periodic elliptic functions that solve the same Yang-Mills equation with maximal O(4) symmetry, with a tunneling period fixed by a parameter $k$~\cite{LS,EXPLO1,EXPLO2}. For $k=1$ the period is infinite and the solution is an instanton with Chern-Simons or topological charge 1, and for $k=0$ the period is finite and the solution is a sphaleron with Chern-Simons $\frac 12$~\cite{LIUNOWAK}.

The exact form of this family of solutions and their period will not be necessary for the rest of the paper as only the values of the parameters entering the charge (\ref{H2}) for $k=0$ (sphaleron path) are needed, i.e.  $\alpha_{0,1,2}(0)\approx (+6, -0.034, +0.165)$.
The topological charge  $q(0)=0$, i.e. the sphaleron carries zero baryon number. It is a boson.
The ratio of the sphaleron mass $M_0$ to the instanton mass $M_1$  is  $M_0/M_1=3\pi/8\sqrt{2}=0.83$. More details regarding this construction
are presented in~\cite{LIUNOWAK}.

The explicit tetraquark states can now be obtained by seeking the eigenstates of (\ref{H1}) for
$k=0$. Specifically, the radial equation for the reduced wavefunction $R_{nl}=u_{nl}/\rho^{\frac 32}$ following from
(\ref{H1}) after inserting (\ref{H2}),  reads ($m_0/m_1=M_0/M_1$)

\be
\label{H5}
-u_{nl}^{\prime\prime}+\frac {g_l(0)}{\rho^2}\,u_{nl}+(m_0\omega_0\rho)^2\,u_{nl}=e_{0, nl}\,u_{nl}
\ee
with the charge $g_l(0)=l(l+2)+2m_0Q(0)$. The energies are
$e_{0, nl}=2m_0(E_{0, nl}-M_0-N_Qm_H)$, with the binding energies as

\be
\Delta_{nl}(0)=E_{0, nl}-N_Qm_H
\ee

The  $1/\rho^2$ potential stems from the kinematical centrifugation plus the repulsion from the U(1)
charge  at the sphaleron point, and is dominant at small distances.

 The parameters $\lambda, m_H,M_1$ are all fixed in the holographic heavy baryon sector with $N_c=3$~\cite{LIUHEAVY}.
A numerical analysis shows that only  for $l=0$, the
$N_Q\leq 3$ states are bound, i.e. open-flavor  tetraquark $QQ\bar q\bar q$.
The S-wave tetraquark  states $QQ\bar q\bar q$ carry
$IJ=00, 01$ assignments  and are degenerate. Heavier exotics are  discussed more thoroughly in~\cite{LIUNOWAK}.

\begin{table}[h]
\caption{Binding energies for tetraquarks versus the $^\prime$t Hooft coupling $\lambda=g^2_{\rm YM}N_c$ with  $M_{\rm KK}=1$ GeV}
\begin{center}
\begin{tabular}{|c|c|c|c|c|}
\hline
$\lambda$ & $QQ\bar q\bar q$ GeV  &  $bb\bar q\bar q $ GeV  & $bc\bar q\bar q $ GeV & $cc\bar q\bar q $ GeV  \\
\hline
10 &$-0.097$ & $-0.088$   & $-0.080$&  $-0.072$\\
15 & $-0.107$  & $-0.091$ &  $-0.077 $&$ -0.062$\\
20&  $-0.108$& $-0.085$  &$-0.064$ &$ -0.041$ \\
25&  $-0.103$& $-0.073$  & $-0.045$ & $-0.018$ \\
30&  $-0.093$& $-0.056$ &$-0.024$& $-0.0016$ \\
32&  $-0.089$& $-0.048$ &$-0.015$ &$0.00073$ \\
\hline
\end{tabular}
\end{center}
\label{tab_bindtet}
\end{table}%

\begin{table}[h]
\caption{Binding energies for tetraquarks versus the $^\prime$t Hooft coupling $\lambda=g^2_{\rm YM}N_c$ with  $M_{\rm KK}=0.475$ GeV. }
\begin{center}
\begin{tabular}{|c|c|c|c|c|}
\hline
$\lambda$ & $QQ\bar q\bar q$ GeV  &  $bb\bar q\bar q $ GeV  & $bc\bar q\bar q $ GeV & $cc\bar q\bar q $ GeV  \\
\hline
10 &$-0.046$ & $-0.044$   & $-0.042$&  $-0.040$\\
15 & $-0.051$  & $-0.047$ &  $-0.044 $&$ -0.040$\\
20&  $-0.051$& $-0.046$  &$-0.040$ &$ -0.035$ \\
25&  $-0.049$& $-0.042$  & $-0.035$ & $-0.028$ \\
30&  $-0.045$& $-0.035$ &$-0.027$& $-0.019$ \\
40&  $-0.031$& $-0.018$ &$-0.0076$ &$0.0011$ \\
\hline
\end{tabular}
\end{center}
\label{tab_bindtetX}
\end{table}%

\section{Efimov states}

  For small distances and S-waves, (\ref{H5}) reduces to

\be
\label{H7}
-u_{n0}^{\prime\prime}+\frac {g_0(0)}{\rho^2}\,u_{n0} \approx e_{0, n0}\,u_{n0}
\ee
For $g_0(0)+\frac 14<0$, the potential in (\ref{H7}) is singular but attractive and leads a priori to
infinitely many bound states, due to the conformal or scale invariance.  The quantization condition converts this continuous symmetry into a
discrete one - the states accumulate at the rate

\be
\label{H10}
\frac{e_{0, (n+1)0}}{e_{0,n0}}=e^{-\frac {2\pi}{\nu_0}}
\ee
with  $\nu_0=\sqrt{-1/4-g_0(0)}$.
This is the essence of the Efimov phenomenon~\cite{EFIMOV,NAIDON}.

\begin{figure}[t]
{%
  \includegraphics[height=4cm]{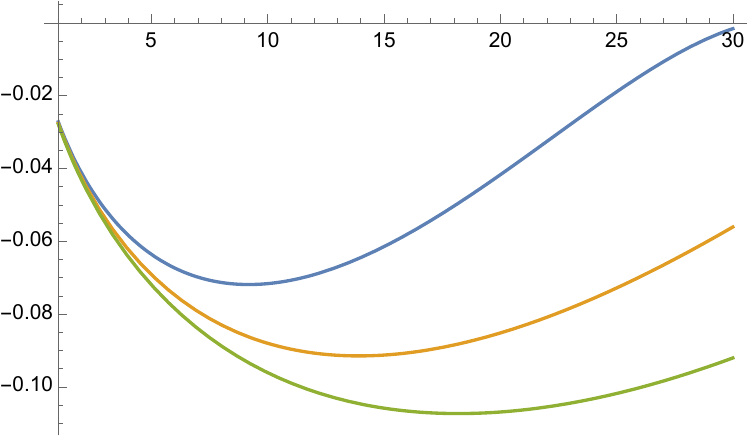}%
}
\caption{Tetraquarks binding energies  in GeV as a function of the $^\prime$t Hooft coupling $\lambda$ for $M_{\rm KK}=1$ GeV:
The upper-blue curve  is for $T_{cc}$, the middle-orange curve  is for $T_{bb}$ and the lower-green curve  is for infinitely heavy quarks.}
\label{fig-data1}
\end{figure}

\begin{figure}[t]
{%
  \includegraphics[height=4cm]{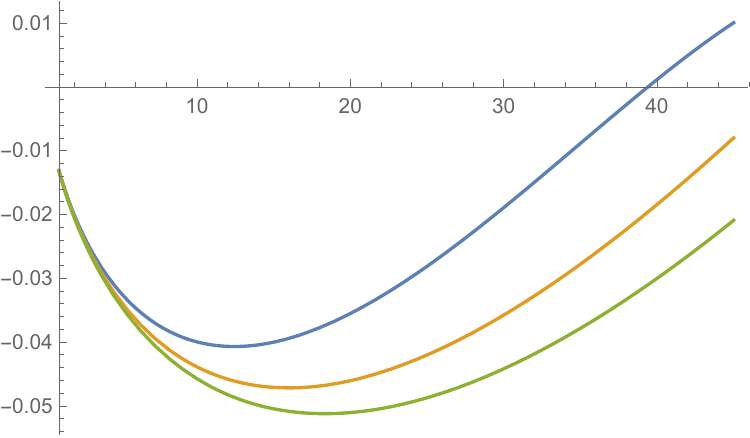}%
}
\caption{Tetraquarks binding energies  in GeV as a function of the $^\prime$t Hooft coupling $\lambda$ for $M_{\rm KK}=0.475$ GeV:
The upper-blue curve  is for $T_{cc}$, the middle-orange curve  is for $T_{bb}$ and the lower-green curve  is for infinitely heavy quarks.}
\label{fig-data2}
\end{figure}

Historically, the Efimov effect originates
from the Borromean  effect, which allows binding of a three-body state even though the two-body state is unbound.  The Efimov equation is  usually
 written in configuration space, and the binding depends on the sign of the  potential $V(R)=\frac{1}{R^2}(s_n^2-1/4)$, where $R$ is a  hyperspherical coordinate stemming from the Jacobi variables for the three-body problem~\cite{NAIDON} (see Eq. 2.32).
 It is remarkable that a similar equation  appears
in a holographic description of an exotic hadron, especially that the physical origin of the $1/\rho^2$ term is different - here it comes from the U(1) Coulomb law  in 1+4 dimensions. For the details of the renormalization of the equation  for the Efimov states we refer to \cite{LIUNOWAK}, and here we only state the main results.

Numerically, the minimal value  $\nu_0\approx \frac 65$ occurs on the sphaleron path, for $N_c=3$,
$N_Q=2$ and $m_H\rightarrow \infty$.
The binding energies for $QQ\bar q\bar q$ depend on the strong  $^\prime$t Hooft coupling $\lambda$ as listed in Table~\ref{tab_bindtet}
for $M_{KK}=1$ GeV~\cite{LIUNOWAK}, and in Table~\ref{tab_bindtetX}  for $M_{KK}=0.475\,{\rm GeV}$~\cite{Liu:2021tpq,Liu:2021ixf}.
The explicit dependence of the binding energy versus $\lambda$ is given
in Figs~\ref{fig-data1}-\ref{fig-data2}, for both values of considered values of $M_{KK}$, respectively.  For both values of $M_{KK}$ the binding energy of the charm tetraquark is few MeV for $\lambda\sim 30$.
Since $e^{-2\pi/\nu_0}\approx 10^{-3}$,   (\ref{H10}) shows that the radially
excited  states rapidly unbind. The  leading
$\lambda /m_H$ heavy mass correction in (\ref{H2}) is repulsive, and penalizes the binding of $cc\bar q\bar q$ 
more than $bb\bar q\bar q$.

Theoretical predictions for the charmed tetraquark $cc\bar{qq}$ are not concise - they vary from binding at the level of 200-300 MeV to unbinding with 200-300 MeV surplus.
Lattice and phenomenological estimates suggest that the double-bottom tetraquark state is deeply
bound with $\Delta_{BB}=-(0.15-0.2)$ GeV~\cite{MALT} (lattice) and $\Delta_{BB}=-(0.17)$ GeV~\cite{KR}
(quark model). The same lattice analysis suggests that the mixed charm-bottom tetraquark state is bound
$\Delta_{CB}=-(0.061-0.015)$ GeV, but the double-charm tetraquark state is not~\cite{MALT}.  Our holographic
results support doubly charmed state, provided we fix the value of 't Hooft coupling at $\lambda \sim 30$, and allows to make predictions for  binding for bottom and mixed bottom-charm states (see Tables I and II for precise numbers).

\section{Decay width}

Recently LHCb has reported a narrow tetraquark $T_{cc}^+$ with a quark content $cc\bar u\bar d$ and
isospin-spin-parity assignment $(0 1^+)$~\cite{LHCb:2021auc}. This  is consistent with the holographic prediction
of bound and degenerate charm tetraquark states $(00^+, 01^+)$ in the heavy quark limit and a strong $^\prime$t Hooft coupling.
The empirical binding energy $\Delta_{CC}$ and width $\Gamma_{CC}$ are relatively small and narrow  as illustrated in Fig~\ref{fig-LHCB}, with

\bea
\Delta_{CC}(T_{cc}^+)&=&-360\pm 40\,{\rm KeV}\nonumber\\
\Gamma_{CC}(T_{cc}^+)&=&+48\pm 2\,{\rm KeV}
\eea

The holographic tetraquarks  is a bound heavy-light vector multiplet  $[0^-, 1^-]$ to a flavor sphaleron core  in bulk with $(00^+)$ assignment.
This is the holographic dual to a  molecule composed of heavy-light  $[D, D^*]$ mesons strongly bound by light meson exchanges on the boundary.  The strong decay
mode of this molecule is natural through

\begin{figure}[t]
\includegraphics[height=4cm]{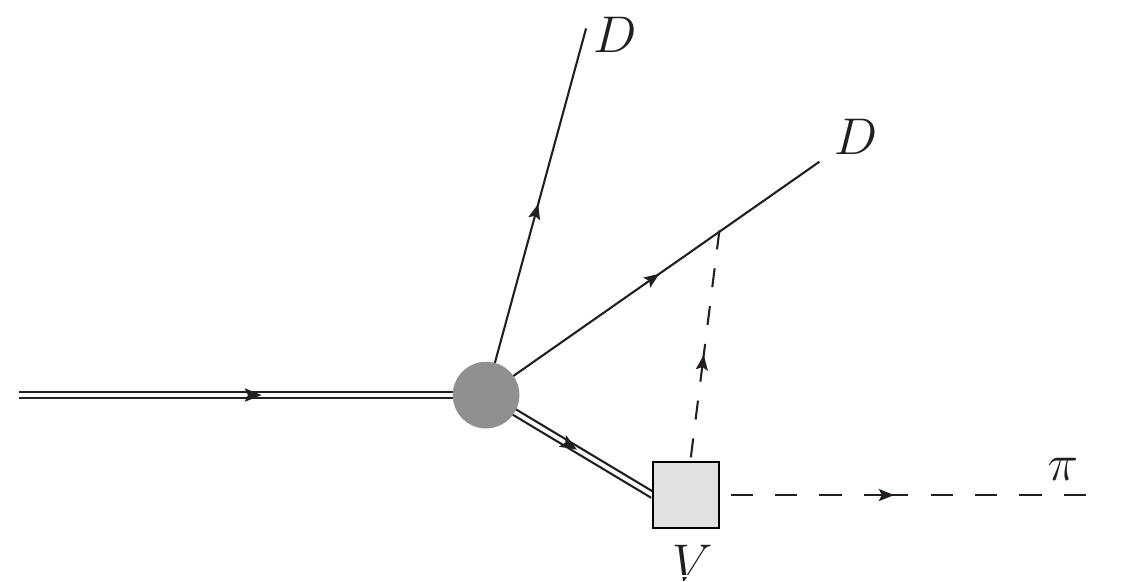}
\caption{The holographic tetraquark $T_{QQ}$ decay process takes place in two steps:
first into ${\rm core}+D+D^{\star}$, then into $D+D+\pi$  when the unstable flavor sphaleron core decays ${\rm core}\rightarrow \pi+\pi$, following
the  recombination $D^{\star}+\pi\rightarrow D$. The incoming-double line denotes the tetraquark $T_{QQ}$, the outgoing double line  the core, the solid lines denote the out-going  heavy-mesons and the dotted lines the  pions. }
\label{fig_diagram}
\end{figure}

\begin{widetext}
\bea
\bigg[T_{cc}(3880)(0 1^+)\bigg]\rightarrow \bigg[{\rm Core}(M_0)[00^+]\bigg]+\bigg[D(1870)\frac 12 0^-\bigg]+\bigg[{D}^*(2010)\frac 12 1^-\bigg]
\eea
\end{widetext}
The unbound isoscalar-scalar flavor sphaleron core $[00^+]$ is unstable and decays subsequently to multi-pions, say minimally to two pions as illustrated in Fig.~\ref{fig_diagram}.  Formally, the decay width is


\begin{align}
\label{GAMMA}
\Gamma_T(S)=\frac{3g^2_H V^2 }{(2S+1)} (2\pi)^4\int d\Phi_3\, |{\cal A}(\vec{k},\vec{p})|^2 \ ,
\end{align}
with $g_H\sim 0.67$, the $D^\dagger \partial_i\pi  D_i$ coupling~\cite{CLEO:2003ggt}.
Here  $S$ is the total spin of the $QQ$ system, and we have used the condition $l=0$. 
We now estimate the  3-body phase space $d\Phi_3$, the decay amplitude ${\cal A}$, and the coupling 
$V$ to  the sphaleron {\rm core} $[00^+]\rightarrow \pi\pi$.

The first part of the decay process $T_{QQ}\rightarrow [00^+]+D+D^{\star}$ is described by  the bulk Chern-Simons term~\cite{Liu:2021ixf}
\bea
\label{CSX1}
    -\frac{i}{16\pi^2}\epsilon_{MNPQ}\Phi^{\dagger}_M\Phi_N\Phi^{\dagger}_P\partial_t \Phi_Q+ {\rm c.c.} \ ,
\eea
in leading order in $\lambda$ and $m_H$.
It is of order  $1/m_H$ since $\Phi_M\rightarrow \Phi_M/\sqrt{m_H}$,  and $\partial_t\Phi_M\rightarrow -im_H\Phi_M$ in the heavy quark limit~\cite{LIUHEAVY}.
The remaining  contributions are suppressed by $1/m_H^2$.
In (\ref{CSX1}) two of the heavy-light fields are valued in the moduli, while the remaining two fields give rise to the decaying heavy-light mesons.
More specifically, using the results in~\cite{LIUHEAVY}, (\ref{CSX1}) gives

\begin{widetext}
\begin{align}
\label{CSX2}
\frac{m_H}{8\pi^2}\left(\frac{1}{\sqrt{16m_Ha N_c}}\right)^2 \phi^2_n(Z) f_{k=0}^2(X-\bar X,Z-\bar Z)
u_{Q,s}^{\dagger}\vec{\sigma}\epsilon \cdot \left(u_{Q,s'}^{\dagger}\vec{\sigma}\times \vec{p}\tau^a \epsilon'\right)a_{Q,s}^{\dagger}a_{Q,s'}^{\dagger}+c.c \ ,
\end{align}
\end{widetext}
with $a=1/(216\pi^3)$.
Here $\phi_n(Z)$ is the wave function for the heavy-light mesons~\cite{LIUHEAVY}

\begin{align}
\phi_n(Z)=\frac{1}{\sqrt{2\kappa}}\frac{1}{\sqrt{2^nn!}}\left(\frac{\sqrt{2}m_H}{2\pi}\right)^{\frac{1}{4}}e^{-\frac{\sqrt{2}\tilde Z^2}{4}}H_n(\tilde Z)
\end{align}
with $\tilde Z=\sqrt{m_H}Z$, and

\begin{align}
f_{k=0}^2(X,Z)
=\frac{\rho^3}{(X^2+ Z^2)^{\frac{3}{2}}}\ ,
\end{align}
the profile function of the fermionic zero mode at the sphaleron point with $k=0$. In  (\ref{CSX2}), the collective sphaleron positions are $(\bar X, \bar Z)$,
$\epsilon$ and $\epsilon'$ are the polarization-isospin vectors for the heavy-light
doublet $(0^-,1^-)$, and $a^{\dagger}_{Q,s}$ is the creation-operator for the heavy-quarks in the bound state.

The second part of the decay process stems from the produced
sphaleron core which is unstable once the heavy quarks are released (recall that stability follows from heavy quark binding).
As a result the isoscalar-scalar flavor core decays through
$[00^+]\rightarrow \pi+\pi$, for a core mass $M_0/M_1=3\pi/8\sqrt{2}=0.83$. This decay is captured by the standard chiral Lagrangian at the boundary, with the sphaleron
core with $k=0$,  described by the monodromy along the holographic direction~\cite{SSX}

\bea
U_{k=0}(x)=e^{i\int_0^\infty dz\,A_z(\vec x, z; k=0)}=e^{i\pi\vec\tau\cdot\hat{x}}\nonumber\\
\eea
The minimal $\pi\pi$-coupling to the monodromy is through the chiral symmetry breaking term in the standard chiral Lagrangian, with $V_{\pi\pi}\sim m_\pi^2$,
hence  $V\sim \sqrt{2M_0} \,V_{\pi\pi}$.
This is an estimate, since the breaking of chiral symmetry in bulk is expected to modify the monodromy for the sphaleron
at the boundary, away from the chiral limit
(much like for the instanton at  $k=1$~\cite{SSX}).

Finally, combining the two decay processes, and using the identity

\begin{align}
\sum_{M}|u_{Q,s}^{\dagger}\vec{\sigma}\epsilon \cdot \left(u_{Q,s'}^{\dagger}\vec{\sigma}\times \vec{a}\tau^a\epsilon'\right)C^{SM}_{ss'}|^2=6|\vec{a}|^2 \ ,
\end{align}
with the spin-isospin modular wavefunctions $T_l(a_4, \vec a)$ defined in~\cite{LIUNOWAK},
the spin-isospin averaged squared  decay amplitude  for the total process shown in~Fig.~\ref{fig_diagram}, reads

\begin{widetext}
\bea\label{eq:amplitudesquare}
|{\cal M}(T_{QQ}\rightarrow D(k)+D(k')+\pi(p))|^2=\frac{g^2_H V^2 }{2(2S+1)}\times 12M_T\times |{\cal A}(\vec k, \vec p)|^2
\eea
with

\begin{align}
&{\cal A}(\vec{k},\vec{p})=\int \frac{d^4p'}{(2\pi)^4}\frac{F(\vec p'+\vec{p})}{128\pi^2 aN_c } \frac{\vec{p}'}{(k-p')^2-m_{D*}^2+i0} \frac{1}{(p^\prime)^2-m_\pi^2+i0}\equiv \frac{\lambda \sqrt{\pi }}{64\pi \kappa}\langle N_Q=2|\rho^3|N_Q=0\rangle  \nonumber \\
&\times  \int \frac{d^3\vec{p}'}{4(2\pi)^3 E_{\vec{p}'}E_{\vec{k}-\vec{p}'}}e^{\frac{1}{8} \sqrt{\frac{3}{2}} (\vec{p}+\vec{p}')^2} K_0\left(\frac{1}{8} \sqrt{\frac{3}{2}} (\vec{p}+\vec{p}')^2\right)\left(\frac{\vec{p}'}{E_{\vec{k}}-E_{\vec{p}'}-E_{\vec{k}-\vec{p}'}}-\frac{\vec{p}'}{E_{\vec{k}}+E_{\vec{p}'}+E_{\vec{k}-\vec{p}'}}\right) \ ,
\end{align}
\end{widetext}
and $\kappa=\lambda a N_c$ in units of $M_{KK}$.
Here $K_0(x)$ is a modified Bessel function.
The momentum dependence of the transition form factor is independent of the modular wave function in the $\rho$-direction. Also the formfactor decays as ${1}/{|p|}$ at large $p$, which is sufficient for the integral to converge, both at large $\vec{p}'$ and $\vec{p}'=-\vec{p}$. The ensuing decay width $\Gamma_T$ is given by (\ref{GAMMA}),
with  the three-body phase space measure

\begin{align}
d\Phi_3=\frac{d^3\vec{p}}{(2\pi)^3 2E_{\vec{p}}}\frac{d^3\vec{k}}{(2\pi)^3 2E_{\vec{k}}}\frac{\delta(M_T-E_{\vec{k}}-E_{\vec{p}}-E_{\vec{k}+\vec{p}})}{(2\pi)^3 2E_{\vec{k}+\vec{p}}}
\end{align}

For an estimate  of the amplitude, let us consider the special point in the phase space where the pion in the final state is at rest, or $\vec{p}=0$.
In this case the momentum of the $D$ mesons are opposite in direction and equal in absolute value: $|k|=|k'|= 8.20$ MeV, assuming that the tetraquark mass $M_T=3880$ MeV and the pion mass is $m_\pi=139$ MeV. Furthermore, since the mass of the $D^{\star}$ meson is much larger than the pion and the final state momentum, and that the pion mass is also much lager than the final state momentum, the momentum $\vec{p}'$ can be approximated by $\frac{m_{D*} }{m_{D*}+m_\pi} \vec{k} \approx \vec{k}$.
Therefore,

\begin{widetext}
\bea
 \int \frac{d^3\vec{p}^\prime\,e^{\frac{1}{8} \sqrt{\frac{3}{2}} (\vec{p}')^2} }{4(2\pi)^3 E_{\vec{p}'}E_{\vec{k}-\vec{p}'}}K_0\left(\frac{1}{8} \sqrt{\frac{3}{2}} (\vec{p}')^2\right)\left(\frac{\vec{p}'}{E_{\vec{k}}-E_{\vec{p}'}-E_{\vec{k}-\vec{p}'}}-\frac{\vec{p}'}{E_{\vec{k}}+E_{\vec{p}'}+E_{\vec{k}-\vec{p}'}}\right)
 =-\frac{m_{D*} \,\vec k}{m_{D*}+m_\pi} \frac{6.2}{4(2\pi)^2}\approx -0.4\,\vec{k} \ ,\nonumber\\
\eea
which is about $3.2$ MeV in magnitude. ${\cal A}$ is about constant, and the phase space volume is then generic

\begin{align}
(2\pi)^4\int d\Phi_3=\frac{0.156}{32\pi^3}\,{\text {\rm MeV}}^2
\end{align}
which amounts to the following contributions to the decay width

\bea
\Gamma_T (S)\approx \bigg[\frac{3g_H^2V^2}{2S+1}\bigg]\times \bigg[\langle 2| \rho^3|0\rangle|^2 \bigg]\times \bigg[\frac{0.4\,|\vec{k}|\sqrt{\pi}}{64\pi \kappa/\lambda}\bigg]^2
\times \bigg[\frac{0.156}{32\pi^3}\,{\text {\rm MeV}}^2\bigg]
\eea
The first bracket  is from the decay couplings, the second   bracket originates from the modular transition vertex, the third bracket is from the
loop integral, and the last bracket is from the integration over the final phase space. $\kappa=a\lambda N_c$ is fixed by the nucleon mass $M_1=8\pi^2\kappa\equiv M_N$
in units of $M_{KK}=m_v/\sqrt{0.67}\sim 1$ GeV~\cite{SSX}, where $m_v$ is a mass of the vector meson $\rho$. An estimate for the $\rho^3$-modular transition matrix element is subtle, since the $|0>$ solution of the Efimov equation is singular and depends on the cut-off~\cite{LIUNOWAK}.
With this in mind, we can estimate the modular transition matrix element as

\bea
\langle 2| \rho^3|0\rangle\sim (\xi\,\langle 2| \rho|2\rangle)^3\sim \xi^3\,\bigg(\frac 1{m_0\omega_1}\bigg)^{\frac 32}
\eea
where the last equality is set by the range of the conformal potential in~\cite{LIUNOWAK}. Here $\xi$ sets the range of our estimate,
with $(\lambda m_0/2)/M_1=M_0/M_1=3\pi/8\sqrt{2}=0.83$,  and $\omega_1=1/\sqrt{6}$ also in units of $M_{KK}$.
Hence

\bea
\bigg(\frac{0.4\,|\vec{k}|\sqrt{\pi}}{64\pi \kappa }\bigg)^2\approx 4.9\times 10^{-6}  \qquad
\langle 2|\rho^3 |0\rangle^2=\xi^6\,\lambda^3 \times 3.18 \, { \rm GeV}^{-6} \qquad
V^2=M_0\times 7.68\,10^{-4} \,{\rm GeV}^4
\eea
which amounts to a relatively small width

\begin{align}
\Gamma_T(S)=\frac{3\lambda^5 g_H^2}{2S+1}\times M_0 \times \xi^6\times 1.88\times 10^{-18} \sim   \frac{1\,{\rm KeV}}{2S+1}
\end{align}
\end{widetext}
with $\xi\sim 5$, $M_0/M_N=0.83$, $g_H\sim 0.67$ and  $\lambda\sim 30$. Note that increasing the value of the parameter $\xi$
by $50\%$  increases  the width by an order of magnitude. The main observation is that for any "natural" value of $\xi$, e.g. not exceeding 10, the smallness of the observed width stems mostly from the very small phase space, combined with the additional suppresion of the core due to the chiral limit.

In sum, the final numerical width is very small with $\Gamma_T(S)\approx 1\,{\rm KeV}/(2S+1)$.  Our results should be considered as an estimate, and not
an absolute prediction, taking into account the large sensitivity of the decay width to the modular transition  matrix element,   and the estimate for
the chiral coupling $V$. However, the qualitative smallness of the width is generic in our analysis, and results from the very small available phase-space,  and  the suppression of the core decay constant $V$  in the chiral limit.
The holographic tetraquark spectrum to order $\lambda/m_H$ in (\ref{H2}),
does not discriminate between the intrinsic heavy quark spin $S=0,1$, making the  tetraquark assignments  $(00^+, 01^+)$  degenerate
by heavy quark symmetry. The degeneracy is lifted by spin-orbit interactions as in~\cite{Liu:2021tpq} for pentaquarks.

However,  we note that in our original work on heavy-baryons, the zero-mode moduli of the heavy-meson field was quantized as a fermion, mostly due to the fact that the leading order Lagrangian in ${1}/{m_H}$ is linear in time-derivative. On the other hand, at sub-leading order, there are quadratic terms in  time derivatives, supporting an alternative quantization as a boson. In  Appendix~\ref{ALTER}, we show how this is implemented for $k=1$ (instanton point), as it carries verbatim for $k=0$ (sphaleron point).  Most notably,  the tetraquark spectrum remains unchanged at quadratic order (the equation of motion remains the same linear equation in the two cases). The decay width assessment is also unchanged at this order. However,  the quantization of the $\chi$-moduli as a {\it boson},  eliminates the $(00^+)$ tetraquark state (intrinsic  spin $S=0$ antisymmetric state),  leaving only a single and non-degenerate $(01^+)$  tetraquark state (intrinsic spin $S=1$ symmetric state). This alternative quantization scheme within holography, appears to be favored by  the current experimental reporting of a single and non-degenerate $T_{cc}^+$ state by LHCb.

\section{Discussions and conclusions}

We have suggested that a heavy and strongly coupled tetraquark emerges in holography as an Efimov state by binding a heavy meson multiplet $(0^-, 1^-)$ to a sphaleron path in D8-D$\bar 8$, with quantum numbers $(00^+, 01^+)$ (fermionic moduli) or $(01^+)$ (bosonic moduli). For  a charmed tetraquark the small binding appears to be consistent with the recently measured tetraquark mass for a strong $^\prime$t Hooft coupling with $\lambda \sim 30$. However, its decay width is
small,  mostly due to the  smallness of the  available phase space, and the suppression of the remaining core decay constant $V$ to pions  in the chiral limit. The single
and narrow $T_{cc}^+$ state recently reported by LHCb is compatible with the $(01^+)$ holographic pentaquark (bosonic moduli).

In our construction, the  tetraquark binding
mechanism is the holographic dual of the Callan-Klebanov mechanism,  albeit for heavier mesons around a topological configuration with fractional Chern-Simons number.  We  have also found the geometrical analogue of the Savage-Wise "supersymmetry" between a heavy antiquark, and a heavy diquark formulated in quark models.

The Efimov  effect  requires that the modulus of the scattering wave, is  much larger than the range for asymptotically weak or power like decaying  potentials. In real physical systems,   the infinite Efimov series truncates to few terms. Actually, the experimental confirmation of the longer hierarchy of states in the Efimov effect was possible only after the discovery of artificial quantum systems  on optical lattices, where one can  control the range and scattering length  through external parameters~\cite{OPTICAL}.
The Efimov "window" in our case, is very narrow too. It is limited by the size of the heavy meson Compton wavelength in relation to the bound state width controlled by  the binding energy. The exponential penalty factor suggests {at most} two bound states, and most probably one, with a typical binding of order few MeV for charm tetraquarks
for $\lambda\sim 30$.

Our holographic  tetraquarks are   different from  the molecules  mediated by pion exchange (deuson with zero heavy flavor) or baryon-antibaryon  states (baryonium) and if also discovered in the bottom sector, will provide the first evidence of a non-conventional, strongly bound cluster different either from a standard meson or a baryon.  Our conclusion is in line with similar recent claims~\cite{KR},  but  the present description is less restrictive (comparing to the quark models)  when it comes to the spin and parity assignment.  The reason is that in our case, the fused heavy quarks are still very strongly correlated with the light flavor degrees of freedom. Needless to say that the holographic construction is  predictive and therefore falsifiable.

\vskip 0.5cm

{\bf Acknowledgements\,\,}
This work was supported by the U.S. Department of Energy under Contract No.
DE-FG-88ER40388 and by the Polish National Centre of Science Grant
UMO-2017/27/B/ST2/01139.

\appendix

\section{Bosonic quantization of the moduli}\label{ALTER}

In this appendix we consider an alternative  quantization of the $\chi$-moduli for $k=1$ (instanton point). 
The same reasoning holds for $k=0$ (sphaleron point).  To order $1/m_H$, the full Lagrangian for the quadratic $\chi$-moduli,
can be read from Eqs. A7 and A21 in~\cite{Liu:2021tpq}, 

\begin{widetext}
\begin{align}
\label{APP1}
    {\cal L}=\frac{1}{2m_H}\dot \chi^{\dagger}\dot \chi+i(1+\frac{3}{2m_Hm_y\rho^2})\chi^{\dagger}\dot \chi+\frac{3}{2m_y\rho^2}\chi^{\dagger}\chi
    +\frac{49}{40m_Hm_y^2\rho^4}\chi^{\dagger}\chi+\frac{33i}{40m_Hm_y\rho^2}\chi^{\dagger}\tau^a\chi \chi^a-\frac{37+12\frac{Z^2}{\rho^2}}{192m_H}\chi^{\dagger}\chi \ .
    \nonumber\\
\end{align}
 with all notations defined therein.
By the replacement $\chi \rightarrow e^{im_Ht}\sqrt{m_H}$, (\ref{APP1}) simplifies

\begin{align}
\label{APP2}
     {\cal L}=\frac{1}{2}\dot \chi^{\dagger}\dot \chi+\frac{3i}{2m_y\rho^2}\chi^{\dagger}\dot \
     \chi-\frac{m_H^2}{2}\chi^{\dagger}\chi+\left(\frac{49}{40m_y^2\rho^4}-\frac{37+12\frac{Z^2}{\rho^2}}{192}\right)\chi^{\dagger}\chi
     +\frac{29i}{40m_y\rho^2}\chi^{\dagger}\tau^a\chi \chi^a\ .
\end{align}
\end{widetext}
This can be interpreted as a system of two harmonic oscillators,
 in a $\rho$ dependent background magnetic field, coupled  by the spin-orbit term.  To proceed, we define
\begin{align}
   \chi=\left( \begin{array}{c}
       x_1+iy_1    \\
        x_2+iy_2
    \end{array}\right) \ ,
\end{align}
and (\ref{APP2}) can be re-arranged
\begin{align}
    &{\cal L}=\frac{1}{2}(\dot {\vec{x_1}}^2+\dot {\vec{x_2}}^2)+\frac{3}{2m_y\rho^2}(y_1\dot x_1-x_1\dot y_1+y_2\dot x_2-x_2\dot y_2)\nonumber\\
    &-\frac{m_H^2+\Omega^2(\rho)}{2}(\vec{x_1}^2+\vec{x_2}^2)+\text{\rm Spin-Orbit} \ .
\end{align}
where $\vec{x}_1=(x_1,y_1)$, $\vec{x}_2=(x_2,y_2)$ and
\begin{align}
    \Omega^2(\rho)=-\frac{49}{20m_y^2\rho^4}+\frac{37+12\frac{Z^2}{\rho^2}}{96} \ .
\end{align}
Here we quantize the theory in the Born-Oppenheimer approximation. We  fix the the modular coordinates $y_I$ and $Z$,
 and first quantize the $\vec{x}_1$ and $\vec{x}_2$ coordinates. This can be justified in the heavy quark limit  with  $m_H$ large,
  where $\chi$ is  fast-moving at frequency $m_H$, while the other degrees of freedom are in slow motion  with a typical
  frequency $\omega_y=\frac{1}{\sqrt{6}}M_{KK}$.

For the $l=0$ state, the spin-orbit coupling vanishes.  
In this case $\vec{x}_1$ and $\vec{x}_2$ decouple,  and we have two identical harmonic oscillators in the background field
\begin{align}
    \vec{A}=\omega_c(y,-x), \omega_c=\frac{3}{2m_y\rho^2} \ .
\end{align}
The spectrum of this system is readily found

\begin{align}
    E=(n_++\frac{1}{2})\Omega_++(n_-+\frac{1}{2})\Omega_- \ ,
\end{align}
with
\begin{align}
\Omega_{\pm}=\sqrt{m_H^2+\Omega^2+\omega_c^2}\pm \omega_c \ .
\end{align}
At large $m_H$, one has
\begin{align}
    \Omega_{\pm}=m_H \pm \omega_c+\frac{\Omega^2(\rho)+\omega_c^2}{2m_H}+{\cal O}\left(\frac{1}{m_H^2}\right) \ .
\end{align}
Clearly, the $\pm$ solutions can be interpreted as {\it bosonic particle and antiparticles}. To leading order in $1/m_H$, 
the two frequencies are identical to the energies following from the fermionic quantization of $\chi$. To be more explicit, one has
\begin{align}
    \Omega_{\pm}-m_H=\pm \frac{3}{2m_y\rho^2}-\frac{1}{10m_Hm_y^2\rho^4}+\frac{37+\frac{Z^2}{\rho^2}}{192m_H}
\end{align}
and the leading order result $\frac{3}{2m_y\rho^2}$ is simply the coefficient  of  the quadratic term of the fermionic Lagrangian.

 \vfil

\end{document}